# BridgeNet: A Dataset of Graph-based Bridge Structural Models for Machine Learning Applications

Bleker, L.[1,2], Güneş, M.C.[1,2], and D'Acunto, P.[1,2,3]

[1]Technical University of Munich, Professorship of Structural Design, School of Engineering and Design, Arcisstraße 21, Munich, 80333, Germany

[2]Technical University of Munich, Munich Data Science Institute, Walther-von-Dyck-Straße 10, Garching, 85748, Germany

[3]Technical University of Munich, Institute for Advanced Study, Lichtenbergstraße 2a, Garching, 85748, Germany

lazlo.bleker@tum.de

**Abstract:** Machine learning (ML) is increasingly used in structural engineering and design, yet its broader adoption is hampered by the lack of openly accessible datasets of structural systems. We introduce BridgeNet, a publicly available graph-based dataset of 20,000 form-found bridge structures aimed at enabling Graph ML and multi-modal learning in the context of conceptual structural design. Each datapoint consists of (i) a pin-jointed equilibrium wireframe model generated with the Combinatorial Equilibrium Modeling (CEM) form-finding method, (ii) a volumetric 3D mesh obtained through force-informed materialization, and (iii) rendered images from two canonical camera angles. The resulting dataset is modality-rich and application-agnostic, supporting tasks such as CEM-specific edge classification and parameter inference, surrogate modeling of form-finding, cross-modal reconstruction between graphs, meshes and images, and generative structural design. BridgeNet addresses a key bottleneck in data-driven applications for structural engineering and design by providing a dataset that facilitates the development of new ML-based approaches for equilibrium bridge structures.

## 1. Introduction

Machine Learning (ML) is increasingly being adopted in the Architecture, Engineering and Construction (AEC) industry, including structural engineering and structural design. Among the various data representations used, graphs stand out as a particularly flexible and powerful medium for ML applications, naturally encoding discrete wireframe structures such as pin-jointed frameworks, discrete form-found structures, and structural frames. Correspondingly, Graph Machine Learning (Graph ML) methods, such as Graph Neural Networks (GNN), can operate directly on graph-structured data and have already been applied to a wide range of problems in structural engineering and design.

Bridge structures stand out as a particularly relevant use case for Graph ML applications: From a structural design perspective, bridges represent a suitable case study for cross-typological design due to their wide variety in topology and geometry of viable designs. From a structural engineering perspective, the substantial forces involved in bridge systems make them compelling candidates for advanced structural analysis and optimization. Moreover, the worldwide aging of existing bridge infrastructure underscores the urgency of developing efficient, data-driven methods for their assessment, rehabilitation, and new design.

## 1.1. Problem Statement

Despite the widely recognized potential of ML applications in structural engineering and design, their adoption is not yet widespread. One of the factors inhibiting the further development of ML in these fields is the lack of generally available training data. At the same time, the most widely applied ML archetype, i.e., supervised learning, is critically dependent on high-quality training data to be able to make accurate predictions. Architecture and engineering firms are understandably reluctant to publicly share the data they have from their design portfolios. On the other hand, the data that is developed and used in academic research is also not always made publicly available. In general, preparing data so that it is accessible to others requires an additional investment of time and resources that may not always be available or prioritized. Instead, most ML models in structural engineering and design are trained on datasets that were specifically developed for the application of interest. This leads to duplicated efforts, hinders reproducibility, and complicates the making of meaningful comparisons across methods. The field, therefore, lacks broadly recognized, openly accessible datasets of wireframe structures that are neutral with respect to specific downstream applications, and rich enough to support a diverse range of data-driven tasks.

## 1.2. Objectives and Scope

This paper introduces *BridgeNet*, a publicly available graph-based dataset of bridge structures intended for ML applications in structural engineering and design. Our focus is on pin-jointed bar structures that satisfy equilibrium through axial forces only, corresponding to the typical level of abstraction for the conceptual structural design stage. Structure are generated using a parametric model based on the Combinatorial Equilibrium Modeling (CEM) form-finding method (Ohlbrock and D'Acunto, 2020). We furthermore derive 3D mesh and 2D image representations from the wireframe graph-based structures, such that each datapoint in BridgeNet is represented in three different modalities. In this way, BridgeNet is positioned to facilitate a wide range of multi-modal ML applications for the conceptual design and analysis of bridge structures.

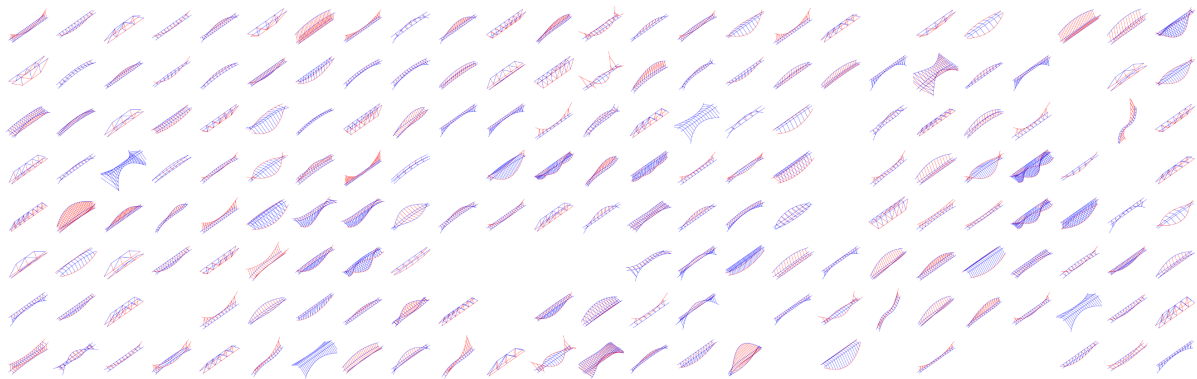

Figure 1. Selection of structures from BridgeNet, visualized with a Self-Organizing Map (SOM).

## 1.3. Contributions

This paper is based on two main contributions:

1. A graph-based dataset of bridge structures (BridgeNet) of 20,000 bridges. For each bridge, the dataset includes: (i) a wireframe model of a pin-jointed structure in equilibrium, (ii) a 3D mesh model based on internal force magnitudes, and (iii) a 2D rendered image of the 3D model from two canonical camera angles. BridgeNet is hosted through Hugging Face and can be accessed at the following link: https://huggingface.co/datasets/lazlo-bleker/bridge-net.

2. A data generation workflow based on form-finding, materialization, and rendering with which BridgeNet is generated, that also serves as a template for the generation of data of additional structural typologies.

## 2. Background and Related Work

### 2.1. Datasets of Structures

While the availability of data for ML applications in structural design and engineering is generally limited, a few publicly available datasets of structures do exist. Particularly in the field of topology optimization, several image-based datasets are available (Sosnovik and Oseledets, 2019; Mazé and Ahmed, 2023; Bastos, 2025; Li et al., 2025). These datasets are typically generated by varying boundary conditions, such as loads and supports, as well as parameters of the optimization process, including the targeted volume fraction. While most of these datasets include 2D structures, Dittmer et al. published a dataset of 3D voxelized topology optimization results (Dittmer et al., 2023).

Topology optimization lends itself well to the creation of automated data generation pipelines, explaining their relative abundance compared to handcrafted or parametric model-based datasets. However, they are also fundamentally limited to the class of optimal structures generated by topology optimization methods, representing only a subset of all structures that are of interest in structural design and engineering. An example of a handcrafted dataset is SimJEB, which contains 3D CAD designs for jet engine brackets crowd-sourced through an open engineering design competition (Whalen et al., 2021). Furthermore, Liao et al. released a dataset of floor plans with annotated shear wall locations collected from architectural design and research institutes (Liao et al, 2021).

Besides volumetric structures, a limited number of wireframe structure datasets have also been made publicly available. Chang and Cheng published a synthetic dataset of rectangular building frame structures (Chang and Cheng, 2020), and Zheng et al. created a dataset of randomized truss structures that occupy a unit cube for metamaterial design (Zheng et al., 2023). Most relevant to this work, Pastrana et al. developed a dataset of CEM topology diagrams of bridge structures (Pastrana et al., 2021). This dataset contains roughly 1600 structures generated to match the geometries of specific bridge typologies, for example, through optimization using automatic differentiation (Pastrana et al., 2022). Unlike the work of Pastrana et al., the 20,000 structures in the dataset presented in this paper are form-found with a specific focus on maximizing their geometrical variety, which is made possible by additional constraints embedded in the CEM algorithm coupled with a data filtering strategy (Section 3.1).

### 2.2. Data-Driven Structure Applications

Despite the limited availability of publicly accessible structural datasets, data-driven applications in structural design and engineering are continuously being developed. What follows is a non-exhaustive list of some ML applications reliant on datasets of discrete wireframe-like structures, such as the one presented in this work.

A long-standing branch of research in this domain concerns *surrogate modeling*, where ML models are aimed at replacing various computationally expensive structural analyses and optimization processes (Mueller, 2014). These include surrogate models for truss Finite Element Method (FEM) analyses (Whalen and Mueller, 2022), cross-section optimization of building frames (Chang and Cheng, 2020), and the amortization of various best-fit form-finding problems (Tam et al., 2020; Tam et al., 2024; Pastrana et al., 2025; Bleker et al., 2025). Datasets of wireframe structures have also supported research on method-specific form-finding problems. For example, the CEM topology diagram dataset by Pastrana et al. was used for the development of a Graph Neural Network (GNN) for CEM edge labeling (Bleker et al., 2022), and later a precursor to the dataset presented in this work was used for training a logic-informed GNN for the same task (Bleker et al., 2024). Finally, ML-based generative models can generate structures similar to those they have been trained on, and therefore benefit greatly from datasets of structures with expansive scopes. This class of models allows new ways of designing structures, such as through latent space exploration of a Variational Auto-Encoder (VAE) (Danhaive et al., 2021), or the generation of designs based on user preferences or text embeddings (Saldana et al., 2021; Guo et al., 2022).

## 3. Dataset Generation

The data generation pipeline used to create BridgeNet consists is based on a sequence of data handling steps (Figure 2). Firstly, equilibrium shapes of bridge structures are generated using a parametric model based on the Combinatorial Equilibrium Modeling (CEM) form-finding method (Ohlbrock and D'Acunto, 2020). Following this, we generate a 3D volumetric mesh of a materialization based on the internal forces of the members of the form-found wireframe structure. Finally, we create image representations of the bridge by rendering the 3D mesh from two canonical camera angles.

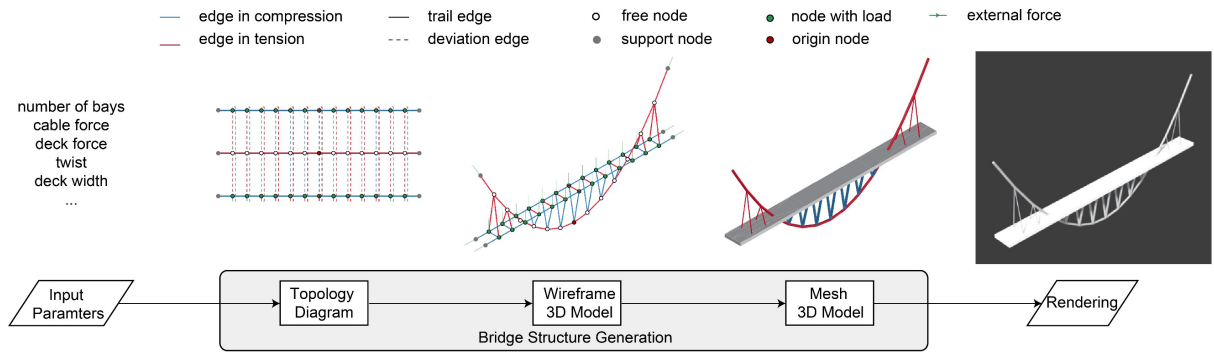

Figure 2. The data generation pipeline used to generate BridgeNet. Bridges are generated using the Combinatorial Equilibrium Modeling (CEM), materialized based on internal force magnitudes and visualized with a render from two canonical camera angles.

### 3.1. Form-Finding with Combinatorial Equilibrium Modeling

The first stage of the BridgeNet data generation pipeline involves creating equilibrium wireframe structures using the Combinatorial Equilibrium Modeling (CEM) form-finding method (Ohlbrock and D'Acunto, 2020). The CEM transforms an input graph, known as the *topology diagram*, into a spatial configuration of nodes and members in a state of static equilibrium. The edges of the topology diagram are separated into *trail edges* and *deviation edges*. Trail edges form ordered chains (trails) that connect an *origin node* to a *support node*. In the case of the bridge structures in BridgeNet, trails are aligned along the length of the bridge, starting with an origin node near the midspan and ending at a support at one of the abutments. Deviation edges constitute the set of remaining edges that connect nodes across two trails. CEM allows metric information regarding basic member lengths and forces to be assigned to trail edges and deviation edges, respectively.

Conventionally, metric parameters, including trail lengths and deviation forces, are assigned a priori. The CEM form-finding algorithm then calculates equilibrium sequentially along the trails of the topology diagram, based on these assigned metric parameters. For the creation of BridgeNet, we use a modified version of the CEM algorithm that allows a higher degree of control over the form-found geometry by automatically assigning selected deviation forces during the form-finding process. Specifically, to maintain control over the curvature and width of the deck, the forces of deviation edges representing *hanger members*, and the forces of deviation edges representing *inter-deck members* (Figure 3) are assigned based on a set of target coordinates for the nodes that constitute the deck of the bridge. The values of these forces are set such that the curvature of the deck corresponds to a predetermined value, and that the width of the deck remains constant along the span of the bridge.

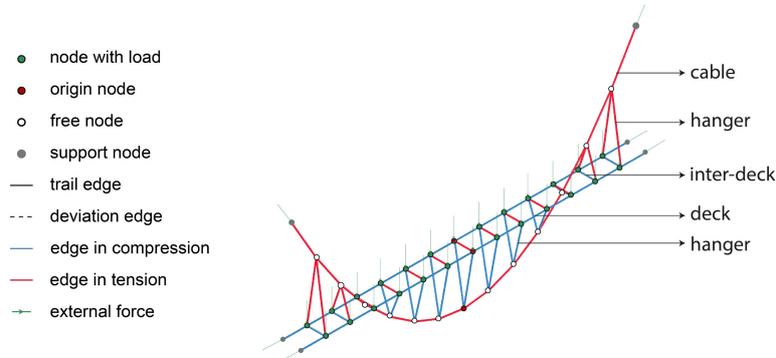

Figure 3. Semantic labels for groups of members of structures in BridgeNet.

We construct a parametric generator based on the CEM that generates topology diagrams from a list of input parameters. These include parameters that determine the topology of the bridge, such as the number of bays. Other parameters define the internal force and geometry of the bridge, such as the force in the main cable and deck, or the degree of twist and width of the deck. To ensure the final dataset only contains bridges that respect reasonable proportions and force distributions, we apply a post-processing filtering step. Specifically, we restrict the generated geometry to a bounding box with horizontal and vertical limits of 0.8 and 0.65 times the span, respectively. Additionally, we discard candidates with an angle between consecutive deck members larger than 15° or a total load path exceeding $0.9 \times$ span$^2$. We generate candidate structures until there are 20,000 structures in the dataset that satisfy all filtering conditions.

### 3.2. Materialization and Rendering

The wireframe geometry form-found with the CEM forms the basis for the subsequent materialization step to obtain a volumetric mesh representation of each structure. The materialization and subsequent rendering steps are based on the methods proposed by Sun et al. (Sun et al, 2024). Specifically, we determine a cross-section size for each member based on its internal axial force. We define circular cross-sections with a diameter ranging between 0.08 meter and 0.8 meter for tension members, and a diameter between 0.25 meter and 1.5 meter for compression elements. For compression members that are part of an arch, we use square tube profiles instead. Finally, the surface spanning between the members of the deck is materialized by a plate with a fixed thickness of 0.4 meter.

Following materialization, we create monochromatic renders of the generated meshes to obtain 2D image representations of each bridge structure. To obtain a neutral visualization, we apply a matte white material to all surfaces in a scene against a dark grey background. For every bridge, a render with a 1200x900 resolution is created from two canonical camera angles, an isometric perspective with a 45° azimuth and 45° elevation, and a side elevation view with a 30° vertical angle.

### 4. BridgeNet

After creation, BridgeNet consists of 20,000 sets of datapoints, each containing a wireframe equilibrium structure, volumetric mesh, and two rendered images. Every datapoint is associated with a model ID and stored inside a folder with the same name (Figure 4). Inside the folder for each datapoint, there is (i) a subfolder containing the rendered images as .jpg files, (ii) a subfolder containing the input parameters of the parametric model used to generate the data as a .json file, (iii) a subfolder containing the wireframe structures as two .json file—one for the node and edge information each—, and (iv) a subfolder containing the volumetric mesh as a .obj file.

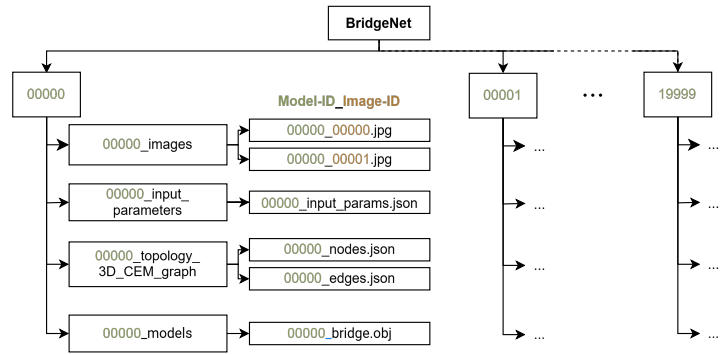

Figure 4. Folder structure of BridgeNet. The data for each bridge is stored inside a folder with its corresponding model ID.

## 4.1. Data Distribution

We report the distributions of topological attributes (number of edges, number of nodes, number of bays), geometrical attributes (deck width, height, mean edge length), and force-related attributes (mean edge force magnitude, ratio of tension edges, total load path) (Figure 5). Some attributes are directly controlled as an input for the parametric model (e.g. deck width) and have a close to uniform distribution, while other parameters have more complex distributions determined by the form-finding and data filtering process.

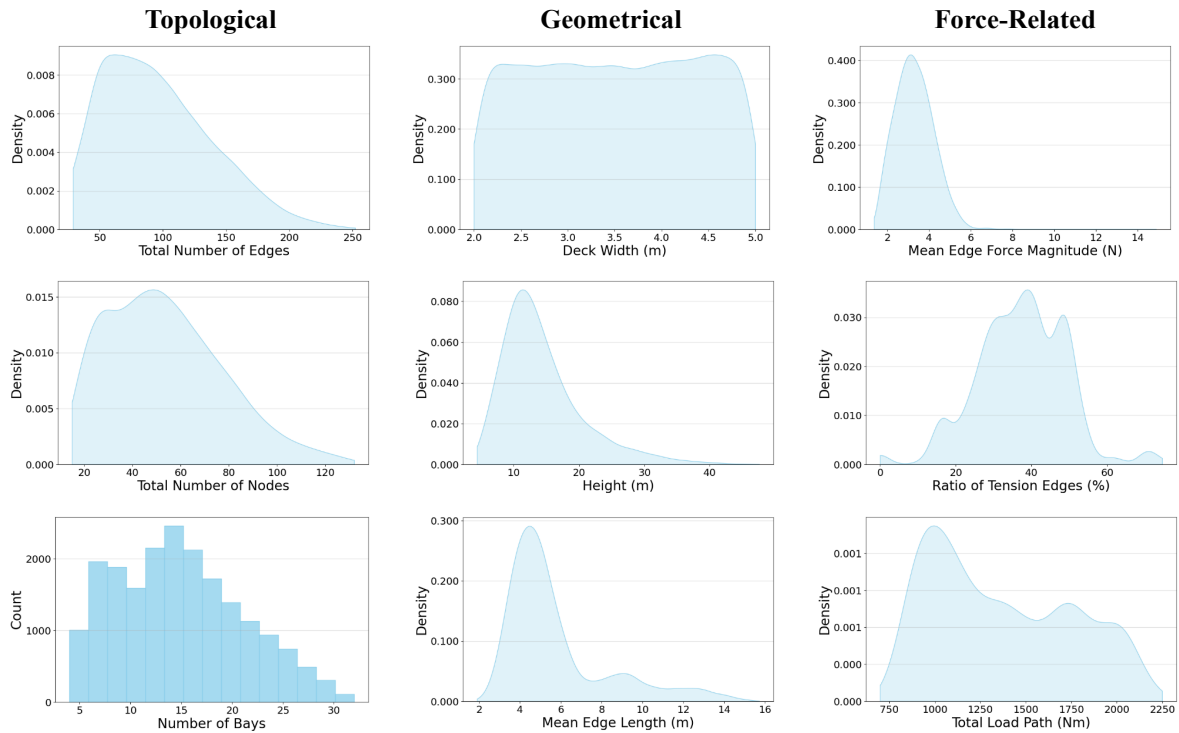

Figure 5: Distributions of parameters and attributes of structures in BridgeNet.

Figure 6 gives an overview of samples of the different bridge typologies in BridgeNet in the three provided data representations.

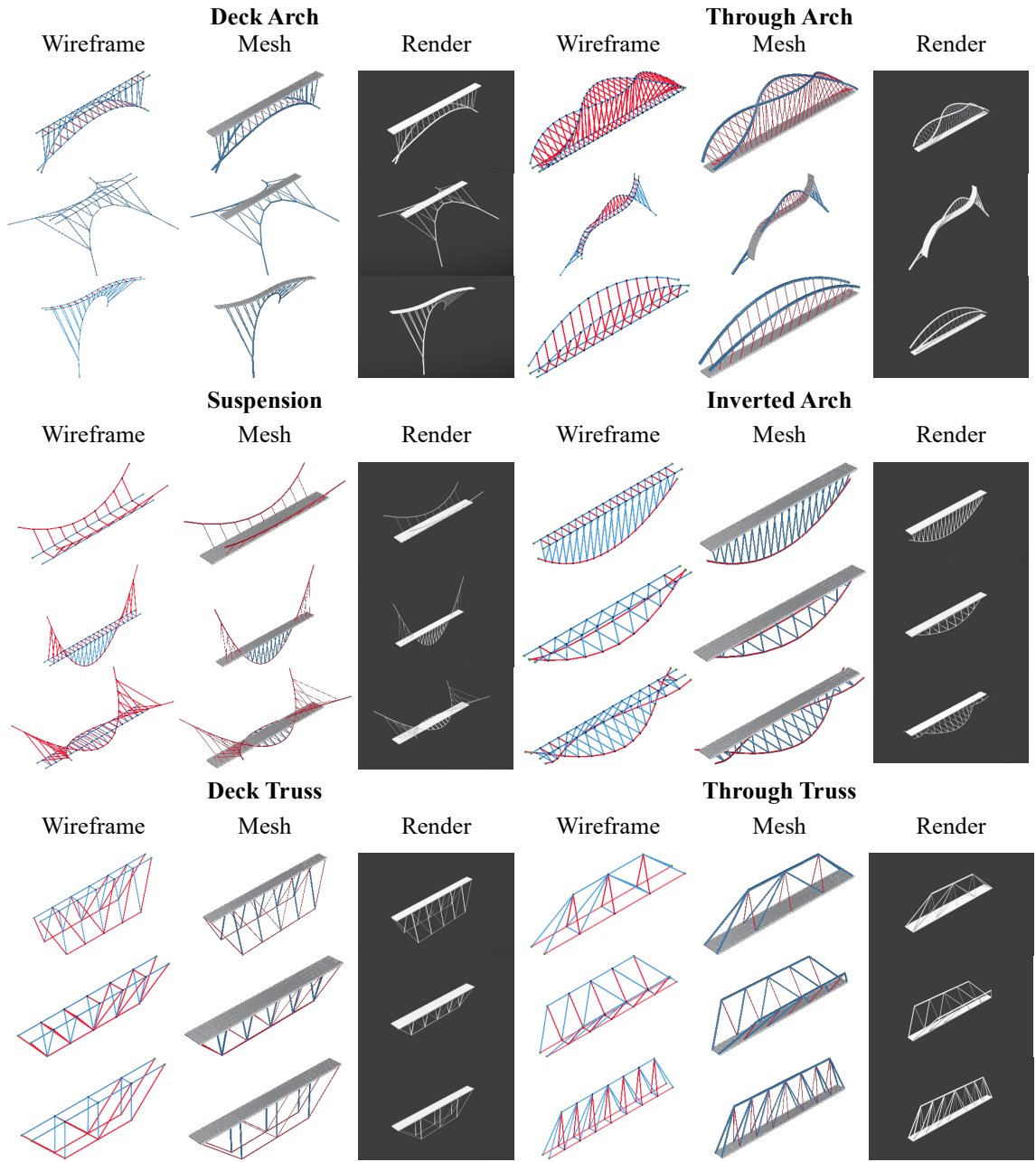

Figure 6: Different bridge typologies represented in BridgeNet.

### 4.2. Tasks

BridgeNet is designed as a flexible, modality-rich dataset that supports a wide spectrum of ML tasks related to structural design and engineering. Owing to its combination of graph-based wireframe data, volumetric meshes, and rendered images, it enables both single-modal and multi-modal learning scenarios. The following categories summarize the key tasks that BridgeNet is intended to support:

**CEM-Specific Tasks.** BridgeNet directly supports tasks related to the Combinatorial Equilibrium Modeling (CEM), including:

- **Edge classification** such as predicting edge labels for trail and deviation edges, as demonstrated on a precursor dataset (Bleker et al., 2024).

- **Parameter inference**, predicting CEM metric inputs (trail lengths, deviation forces, origin node coordinates) from partial data.

- **Forward surrogate modeling**, where ML models approximate the CEM forward form-finding process to reduce computation time.

**Multi-Modal Learning Tasks.** The inclusion of graph, mesh, and image modalities allows models to learn cross-modal relationships:

- **Image-to-wireframe** workflows such as predicting a structural graph from an image.

- **Mesh-to-wireframe** reconstruction for data completion or model-to-model conversions.

- **Cross-modal alignment**, in which representations from graphs, meshes, and images are jointly embedded into a unified latent space. These tasks enable the integration of graph-based models with computer vision or 3D deep learning architectures.

**Generative Structural Design.** The geometric diversity of BridgeNet makes it well suited for generative modeling approaches, such as:

- **Graph-based generative models** for synthesizing new equilibrium structures, either directly through nodal coordinates or indirectly via form-finding input parameters.

- **Parametric generative models**, where ML models propose new parameters as an input for the CEM-based parametric model.

- **Image-conditioned structural generation**, enabling user-guided design exploration based on input images.

## 5. Conclusions

This paper presented BridgeNet, a publicly available dataset of 20,000 form-found bridge structures represented across three modalities: graph-based wireframes, volumetric meshes, and rendered images. By combining CEM form-finding with a unified workflow for materialization and visualization, BridgeNet provides an extensible application-agnostic dataset for ML research in structural engineering and design. In this way, BridgeNet helps to alleviate the prevalent shortage of data in the field, and can form a basis for benchmarking, comparison, and reproducibility across ML methods.

Nevertheless, several limitations remain. The dataset currently focuses exclusively on pin-jointed, axially loaded structures in equilibrium, and therefore does not include stiffness effects, bending behavior, or material nonlinearity. The typological range is still limited to a subset of bridge forms defined by the chosen parametric model. Moreover, all data is synthetically generated, with no real-world bridge geometries or realistic loading scenarios included.

Future work will therefore focus on expanding BridgeNet along three directions: (i) broadening the typological and geometric diversity of generated bridges, (ii) incorporating additional layers of structural information, such as FEM-based analyses, and (iii) integrating real-world bridge data to enable hybrid synthetic-real training and validation. Finally, the presented workflow can be extended to generate datasets of other structural typologies beyond bridges as well.

## 6. Acknowledgments

This work is supported in part by the International Graduate School of Science and Engineering (IGSSE) of the Technical University of Munich (TUM) and in part by the TUM Georg Nemetschek Institute (GNI) under the Artificial Intelligence for the Automated Creation of a Digital Archive of Bridge Infrastructure Project.